\documentclass{PoS}

\title{SuperCDMS SNOLAB Low-Mass Detectors: Ultra-Sensitive Phonon Calorimeters for a Sub-GeV Dark Matter Search}

\ShortTitle{SuperCDMS SNOLAB Low-Mass Detectors}

\author{\speaker{Noah Kurinsky}, Paul Brink, Richard Partridge, Blas Cabrera \\
        Stanford University \& SLAC National Accelerator Laboratory \\
        E-mail: \email{Noah.Kurinsky@stanford.edu}}

\author{Matt Pyle\\
        University of California at Berkeley}
        
\author{On Behalf of the SuperCDMS Collaboration\thanks{http://cdms.berkeley.edu/}}

\abstract{We present the technical design for the SuperCDMS high-voltage, low-mass dark matter detectors, designed to be sensitive to dark matter down to 300 MeV/$c^2$ in mass and resolve individual electron-hole pairs from low-energy scattering events in high-purity Ge and Si crystals. In this paper we discuss some of the studies and technological improvements which have allowed us to design such a sensitive detector, including advances in phonon sensor design and detector simulation. With this design we expect to achieve better than 10 eV (5 eV) phonon energy resolution in our Ge (Si) detectors, and recoil energy resolution below 1eV by exploitng Luke-Neganov phonon generation of charges accelerated in high fields.}

\FullConference{38th International Conference on High Energy Physics\\
		3-10 August 2016\\
		Chicago, USA}

\begin{document}

\begin{figure}[t]
\centering
\includegraphics[width=.43\textwidth,clip=True,trim=5 5 0 5]{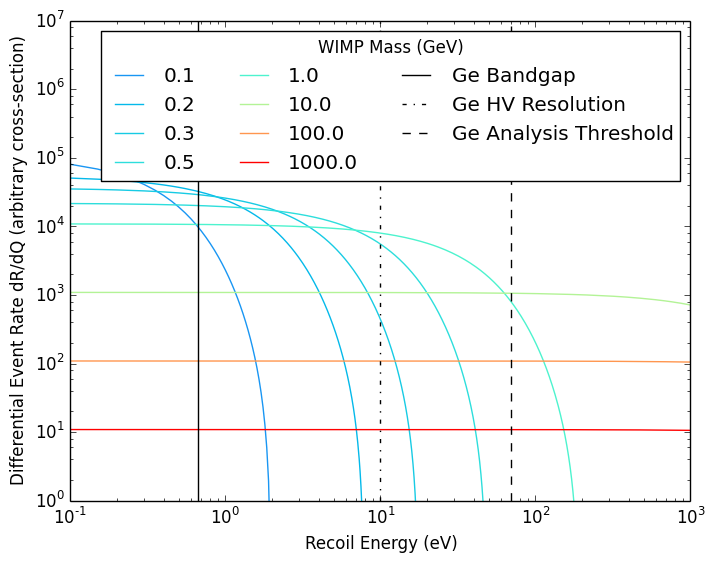}
\includegraphics[width=.43\textwidth,clip=True,trim=5 5 0 5]{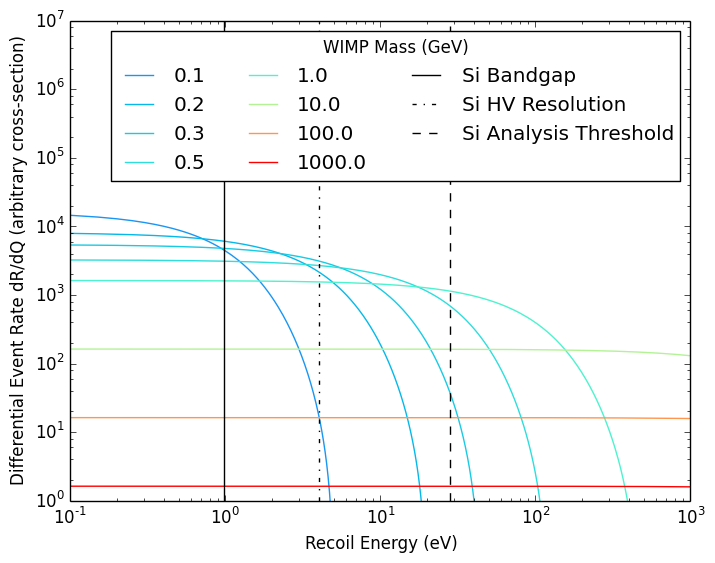}
\caption{Recoil energy spectra for dark matter of various masses for the standard spin-independent WIMP-like coupling, assuming Maxwell-Boltzmann velocity distribution centered on 220 km/s mean velocity. Left is the rate in Ge, right in Si. The event rate is less in Si due to the lower atomic weight for the same number density and nucleon cross-section, but the lighter nucleus means there is less kinematic suppression for Si, and thus lower masses can be probed with the given phonon energy thresholds.}\label{fig:spectrum}
\end{figure}

SuperCDMS SNOLAB is a second-generation dark matter (DM) experiment designed to probe dark matter masses down to 300 MeV/$c^2$ and reach the coherent neutrino scattering floor below 10 GeV/$c^2$\cite{sensitivity}. This is a sensitivity improvement of over three orders of magnitude from the first generation experiment, and is the result of advances in sensor technology, increased target mass, and a quieter experimental environment. Our low-mass reach is driven by `High Voltage' (HV) detectors, which use large electric fields to force charge carriers to excite additional phonons, amplifying the phonon energy of nuclear and electronic recoils. This technique was demonstrated previously with the CDMSlite\cite{cdmslite} analysis, which has world-leading limits for 1-5 GeV/$c^2$. 

The low-mass reach for any direct-detection experiment is kinematically limited by the velocity distribution of the dark matter, nuclear and fiducial mass of the target, and energy resolution of the measurement, which determines the energy threshold. Figure 1 demonstrates this effect, where low mass recoil spectra drop precipitously above the mean velocity peak, and below a certain mass a fixed threshold precludes any detection regardless of fiducial volume. It is this constraint that has motivated us to create detectors with phonon energy resolutions an order of magnitude better than our previous designs, with recoil energy resolution on the order of 100 meV when operated at high bias to further amplify the phonon signal through production of Luke-Neganov phonons\cite{sensitivity}.

\begin{figure}[t]
\centering
\includegraphics[width=.47\textwidth]{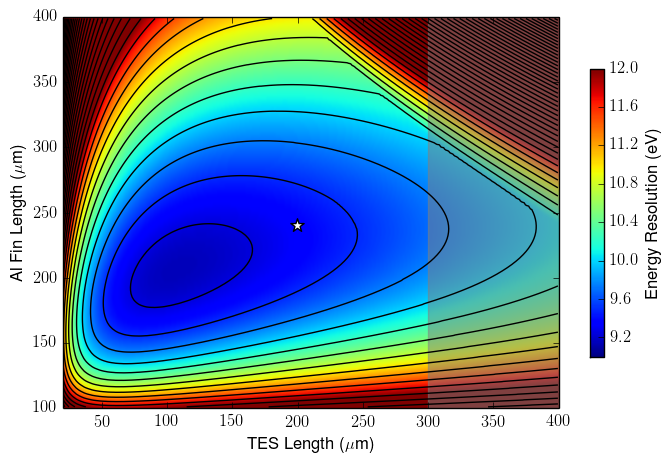}
\includegraphics[width=.47\textwidth]{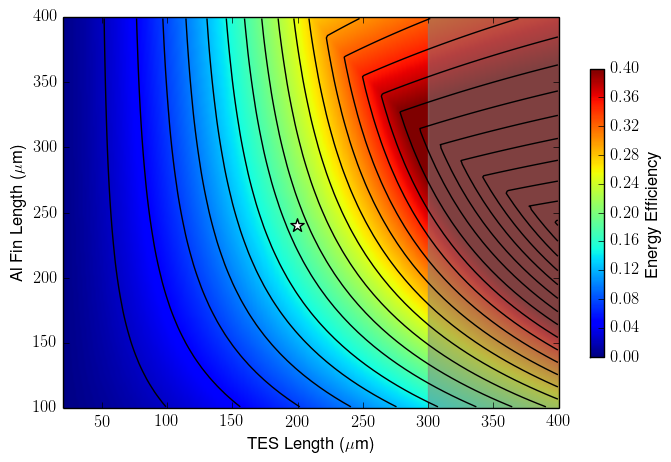}
\caption{Left: Energy resolution versus QET fin length and TES length, with contours spaced every 0.25 eV in energy resolution. Soft optimum shows fairly large freedom within the range 150-300 microns in fin length, and 50-350 microns in TES length, shaded region denotes the phase-separated regime. Right: Collection efficiency for phonons on their first surface contact, while they are still largely position dependent; this is a proxy for the expected fiducialization signal to noise of the detector. The star shows the chosen dimensions based on considerations discussed in the text.}\label{fig:optimization}
\end{figure}

Our phonon sensors consist of aluminum films surrounding tungsten transition edge sensors, tessellating the surface of an ultra-pure Ge or Si crystal, wired in parallel to form 6 readout channels per crystal face. Athermal phonons generated either by the initial interaction, or by electrons and holes being drifted through the crystal, break cooper pairs in the aluminum fins, which are drifted to the aluminum-tungsten interface. At the interface, the proximity effect between aluminum and tungsten reduces the pair energy, effectively trapping the quasiparticles until they recombine, emitting energy both in down-conversion and recombination phonons, which heat the TES. The TES in turn is voltage biased and the current through the TES is measured by a SQUID, to obtain a high signal-to-noise estimate of the TES resistance, and therefore temperature change, allowing a measurement of absorbed phonon energy.

The TES response is dependent on the cooling timescale of the TES, phonon absorption rate, and thermal fluctuation noise between the TES and substrate\cite{pyle}. In addition, the energy resolution is limited by diffusive losses in the aluminum fins\cite{quasiparticles} and energy transport efficiency at the interfaces of the QET\cite{pyle}. Using an optimum filter estimate of the pulse energy, we find that the expected energy resolution of a phonon channel composed of parallel QETs follows the relation\cite{pyle}
\begin{equation}
\sigma_{E}^2 =\frac{2nKk_bT_c^{n+1}}{\epsilon^2}\left(\tau_{pulse}+\frac{2}{n}\tau_{TES}\right)\stackrel{n=5, \tau_{pulse} > \tau_{TES}}{\longrightarrow} \frac{10Kk_bT_c^{6}}{\epsilon^2}\tau_{pulse}
\end{equation}
where $T_c$ is the TES transition temperature, $K$ is the bath thermal conductance, $\epsilon$ is the energy collection efficiency, $\tau_{pulse}\propto f_{al}v_{phonon}/t_{crystal}$ is the phonon absorption time, $f_{al}$ is the fractional aluminum coverage of the crystal, and $\tau_{TES}$ is the response time of the of the voltage-biased TES. 

The largest effect on energy resolution comes from the transition temperature, up to the point where $\tau_{TES}$ becomes dominant, and then from QET optimizations which increase phonon absorption rate and phonon collection efficiency. Figure \ref{fig:optimization} shows the chosen QET parameters compared to calculations for energy resolution, which include the effect of efficiency and aluminum coverage, as well as `first-pass' phonon collection efficiency, which is a proxy for the position dependence of the signal. Both plots assume transition temperature of 45 mK, and use numbers for a Ge substrate. These calculations assume that the TES is in thermal equilibrium, which at 45 mK is true for devices well below about 300 microns in length\cite{pyle}\footnote{Based on preliminary measurements at UCB suggesting devices were phase separated at 220 microns in length, with Tc at 65 mk, and prototypes of this design with Tc$\sim$85 mK.}; the phase separated (non-equilibrium) regime is shown in gray in Figure \ref{fig:optimization}, and was avoided in this design to ensure predictable sensor performance and optimal energy resolution.

\begin{figure}[t]
\centering
\includegraphics[width=.47\textwidth,clip=True,trim=25 10 30 25]{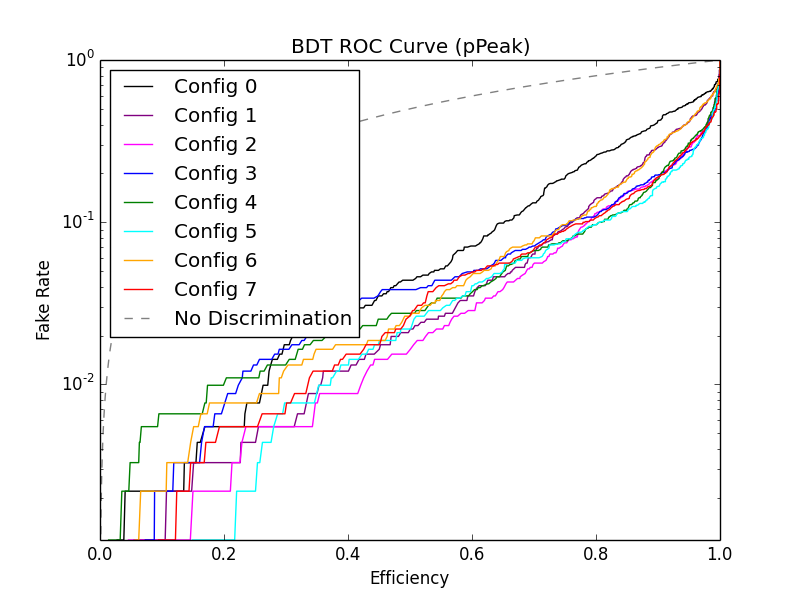}
\includegraphics[width=.47\textwidth,clip=True,trim=25 10 30 25]{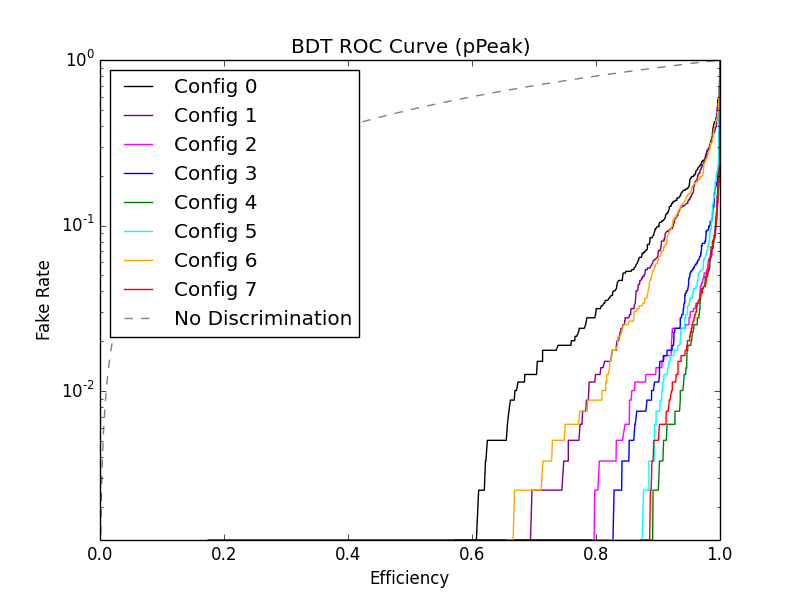}
\caption{BDT rejection curves in log-linear space, with surface acceptance (fake rate) versus signal acceptance. Configuration 4 was chosen for optimal performance at 10 GeV/$c^2$ (right), and good performance at 1GeV/$c^2$ before energy resolution restrictions prevent rejection of proper fiducialization. It should be noted that no analysis threshold is placed on either of these plots, such that they represent worst-case rejection, but rejection improves with increased energy threshold.}\label{fig:bdt}
\end{figure}

Figure \ref{fig:optimization} also shows that the energy resolution optimization allows a large parameter space meeting our 10 eV resolution requirement, and we increased TES length as much as possible while giving the phase separation regime a wide berth. We also increased Al fin length from the strict optimum to increase the position dependence of the phonon signal, in order to improve our ability to reconstruct and reject surface events, the majority of which are electron recoils, which can no longer be removed based on charge yield and surface fields as in the iZIP detectors. By increasing the aluminum coverage from 4\% in the IZIP detectors to 30\% in the HV detectors, we expect our phonon signals to be $\sim$10 times more position dependent (informed by past phonon-only devices). 

In addition to the QET design, the phonon channel layout has a large effect on the fiducialization ability of the detector due to the limited number of channels, only six per face. To choose a sensor layout, we employed the detector Monte Carlo to generate a sample of WIMP-like recoil events, and simulated the detector response for eight different channel configurations deemed robust to various detector failure modes. We compared their performance by partitioning energy into each channel, simulating detector response, adding TES, circuit, SQUID, and environmental noise, and fitting the resulting pulses to obtain optimal-filter estimates of the true TES response. From these pulses we derived a handful of position-dependent metrics, which were used to train and cross-validate a Boosted Decision Tree (BDT) in order to quantify the position reconstruction power of each configuration. The results of this optimization can be seen in Figure \ref{fig:bdt}. The configuration discussed in \cite{sensitivity} was chosen based primarily on the fiducialization performance at 10 GeV/$c^2$, where surface backgrounds will be a larger limitation on sensitivity. The 1 GeV/$c^2$ spectrum is low enough in recoil energy that the predominant backgrounds will be isotropic throughout the crystal, and surface rejection will not impact the science yield nearly to the same extent.

\end{document}